\documentclass[11pt,a4paper]{article}
\pdfoutput=1
\usepackage{jcappub}
\usepackage[T1]{fontenc} 
\usepackage{amsmath,amssymb}

\newcommand{\bk}{{\mathbf k}}

\newcommand{\bn}{{\mathbf n}}

\newcommand{\OO}{{\cal O}}
\newcommand{\JJ}{{\cal J}}

\newcommand{\PP}{{\cal P}}
\newcommand{\HH}{{\cal H}}

\newcommand{\de}{\delta}
\newcommand{\De}{\Delta}

\newcommand{\ga}{\gamma}

\newcommand{\ka}{\kappa}

\newcommand{\La}{\Lambda}

\newcommand{\Om}{\Omega}

\newcommand{\be}{\begin{equation}}
\newcommand{\ee}{\end{equation}}
\newcommand{\gsim}{\stackrel{>}{\sim}}

\newcommand{\bea}{\begin{eqnarray}}
\newcommand{\eea}{\end{eqnarray}}
\newcommand{\bean}{\begin{eqnarray*}}
\newcommand{\eean}{\end{eqnarray*}}

\title{The Cosmological Consistency Relation in a Universe with Structure}

\author{Varun Rustagi}
\author{and Ruth Durrer}

\affiliation{Universit\'e de Gen\`eve, D\'epartement de Physique Th\'eorique and Centre for Astroparticle Physics,
24 quai Ernest-Ansermet, CH-1211 Gen\`eve 4, Switzerland}

\emailAdd{varun.rustagi@etu.unige.ch}
\emailAdd{ruth.durrer@unige.ch}

\abstract{In this short paper we determine the effects of structure on the cosmological consistency relation which is valid in a perfect Friedmann Universe. We show that within $\La$CDM the consistency relation is violated by about 1.5\% for redshifts $z\simeq 2$ and this violation raises up to  
2\% at  $z\simeq 5$ and 3\% at $z\simeq 10$ after which it settles at about 2.7\% for $z>10$. This effect of cosmic structure on the distance redshift relation is also very sensitive to the determination of the dark energy equation of state via cosmic distances. It actually leads to an {apparent} unphysical behavior of  $w(z)$ which even diverges at $z\sim 2.5$ and which we also discuss here.}

\begin{document}

\maketitle

\section{Introduction and motivation}
\label{s:intro}
In a perfectly homogeneous and isotropic Universe, the area distance and the Hubble parameter obey a so called consistency relation~\cite{Clarkson:2007bc,Clarkson:2007pz} which is satisfied independently of the matter content of the Universe or of its spatial curvature. However, the observed Universe is not  perfectly homogeneous and isotropic but contains galaxies, clusters, filaments and voids. The presence of these cosmic structures is expected to modify the consistency relation. In this paper we want to study the extent of this 'back reaction' and its character. In a futuristic high precision test of the consistency relation, the presence of structure might even represent an additional test of $\La$CDM and the relatively late formation of structure in this model.

The curvature parameter in a Friedmann model is defined as
\be
\Om_k = -\frac{K}{a_0^2H_0^2}
\ee
where $K$ is the (dimensionless) spatial curvature, $H_0$ is the present Hubble parameter  and $a_0$ is the present scale factor.
The spatial metric can be given as
\bea
a^2(t)\ga_{ij}dx^idx^i &=& a^2(t)\left(d\chi^2 +S_k^2(\chi)d\Om^2\right) \qquad \mbox{ where}\\
S_k(\chi) &=&\left\{ \begin{array}{ll} \sinh\chi & \quad K=-1 \\ \chi &  \quad  K=0\\ \sin\chi  & \quad  K=1 \,,
\end{array}\right. 
\eea
where $d\Om^2$ denotes the metric of the 2-sphere.
Note that in this form $\chi$ is dimensionless while $a(t)$ has the dimension of a length. Also, having set $K=\pm 1, 0$, except in the case $K=0$,
 we can no longer set $a_0=1$ since $a_0$ is the present curvature radius. In a perfect Friedman universe the area distance is given by ($c=1$)
\be
 d_A(z) = \frac{1}{(1+z)}\frac{\sin\left(\chi a_0H_0\sqrt{-\Om_k}\right)}{H_0\sqrt{-\Om_k}}\,, \qquad
 \chi(z) = \frac{1}{a_0}\int_0^z\frac{dz'}{H(z')}\,.
\ee
This formula is correct also for negative curvature, where $\Om_k>0$ since in this case both, the numerator and the denominator of the above fraction are imaginary. In the case $\Om_k=0$ it has to be interpreted as a limit.

Introducing the comoving distance
$D(z) = \frac{a_0}{a(z)}d_A(z) = (1+z)d_A(z)$ one finds after some algebra~\cite{Clarkson:2007pz}
\be\label{e:cons}
\Om_k = \frac{(D'(z)H(z))^2 -1}{D^2(z)H_0^2} \,.
\ee
The prime denotes a derivative with respect to redshift $z$. Hence in a perfect Friedman universe the right hand side of Eq.~\eqref{e:cons} is independent of redshift, independent of the matter content of the Universe and equal to the present curvature parameter.

In this paper we want to study what happens when taking into account fluctuations in the matter distribution. For this we assume that the quantity $d_A(z)$ is determined as an  average over 
as many measurements as we wish and we replace this by the ensemble average. From first order perturbations we therefore expect no effect on the  averaged $d_A(z)$. However, the angular diameter distance has been determined also to second order in perturbation theory~\cite{BenDayan:2012wi,Fanizza:2013doa,Umeh:2012pn,Umeh:2014ana,Marozzi:2014kua}. At second order, the  average perturbation does not vanish but gives a contribution of several percent to the distance which was determined in~\cite{Clarkson:2014pda}. In this paper it was found that the perturbation to the distance at low redshift, $z\lesssim 0.1$, is dominated by the Doppler term while at higher redshift, $z\gtrsim 0.2$, it is dominated by the lensing contribution. This is also in agreement with the first order results on the fluctuation of the area distance which were determined in Ref.~\cite{Bonvin:2005ps}.

In this paper we concentrate on the lensing term which dominates the result for $z>0.2$. Of course not only the distance but also the redshift is perturbed by inhomogeneities which introduce a Doppler and  
gravitational potential term and more~\cite{Umeh:2012pn,BenDayan:2013gc,Marozzi:2014kua}. However, these perturbations are always much smaller than the lensing term considered here. As they are proportional to the velocity or to the gravitational potential, they are suppressed by factors $\HH/k$ and $(\HH/k)^2$ with respect to the lensing term which is proportional to the Laplacian of the gravitational potential. Here $k$ denotes the wave number of the perturbation while $\HH=aH$ is the conformal Hubble parameter. For this reason we neglect redshift perturbations and use the redshift of the background Friedmann Universe in this work.

The paper is structured as follows: In the next section we present the second order fluctuation of the area distance and compute its expectation value as a function of redshift in a standard spatially flat $\La$CDM Universe. Details of the derivation are deferred to an appendix. The linear and halofit power spectra are computed using {\sc class}~\cite{class1,class2}. In Section~\ref{s:res} we determine the effect of the modified distance-redshift relation on the consistency relation by calculating  the right hand side of~\eqref{e:cons} which vanishes in the input background cosmology. We shall find that the resulting $\Om_k$ is not only non-zero but also redshift dependent. We  then use the first and second derivatives of the distance to determine the equation of state of dark energy, $w(z)$.
The equation of state determined in this way, which is correct only in a perfect Friedmann Universe,  is strongly affected by clustering for $z\gtrsim 1.5$. It not only becomes smaller than $-1$ but it even diverges at $z\simeq 2.5$ and becomes small but positive at higher redshifts.
 In Section~\ref{s:con} we summarize our results and conclude.
\vspace{0.1cm}\\
{\bf Notation}: We set the speed of light, $c=1$ throughout. We use the perturbed Friedmann metric in longitudinal gauge,
\be
 ds^2 = a^2(t)\left[ -(1+2\Psi)dt^2 +(1-2\Phi)\ga_{ij}dx^idx^j\right]\,.
\ee
The gravitational potentials $\Phi$ and $\Psi$ are the Bardeen potentials. In a $\La$CDM cosmology we have $\Phi\simeq\Psi$ to very good accuracy. We denote the derivative by conformal time with a dot so that the Hubble parameter is given by
\be
H =\frac{\dot a}{a^2}= \HH/a \,.
\ee

\section{The effect of structure on cosmological distances}
\label{s:struc}
The area distance in a generic spacetime from an observer to a source on her background lightcone at spacetime in direction $\bn$ and at redshift $z$ is given by
\be
d_A^2(z,\bn) =\det\JJ(z,\bn)  \,,
\ee 
where $\JJ$ is the Jacobi map, see e.g.~\cite{Perlick:2004tq}. The luminosity distance $d_L$ which is the one truly observed e.g. in Type IA supernovae is simply related to $d_A$ by the so called Etherington reciprocity relation~\cite{Perlick:2004tq},
\be
d_L(z,\bn) = (1+z)^2d_A(z,\bn) \,.
\ee
In Ref.~\cite{Clarkson:2014pda} it was shown that for redshifts $z\gsim 0.2$, the largest contribution to the ensemble average of the area distance
at second order in cosmological perturbation theory is given by the lensing term,
\be
\langle d_A(z,\bn)\rangle  = \bar d_A(1+\De(z)) \,, \qquad \De(z) = \frac{3}{2}\langle \ka^2(\bn,z)\rangle  \,,
\ee
where $\ka$ is the convergence in the first order Jacobi map. 

This result can be understood easiest by referring to the conservation of surface brightness by gravitational lensing. As surface brightness is proportional to $d_A^{-2}$ this implies that this quantity has to be conserved order by order. Expanding $d_A$ to second order in perturbations,  $d_A=\bar d_A + \de^{(1)} +\de^{(2)} +\OO(3)$ we have
\bea
\bar d_A^{-2} = d_A^{-2} &=& \frac{1}{\bar d_A^2+2\bar d_A\de^{(1)}+2\bar d_A \de^{(2)}+(\de^{(1)})^2} +\OO(3) \nonumber \\
&=&  \frac{1}{\bar d_A^2}\left[1-2\frac{\de^{(1)}}{\bar d_A}-2\frac{\de^{(2)}}{\bar d_A}+3\left(\frac{\de^{(1)}}{\bar d_A}\right)^2  +\OO(3)\right] \,.
\eea
Taking the ensemble average  and using $\langle \de^{(1)}\rangle  =0$, this implies
\be
\De=\frac{\langle\de^{(2)}\rangle}{\bar d_A} = \frac{3}{2}\frac{\langle(\de^{(1)})^2\rangle}{\bar d^2_A} = \frac{3}{2}\langle\ka^2\rangle \,.
\ee
For the last equal sign we used that the first order perturbation from lensing in the angular diameter distance is given by $\de^{(1)}/\bar d_A= -\ka$.

The Jacobi map to first order in a perturbed Friedmann Universe is~\cite{RuthBook} 
\bea
\JJ_{ab}(\theta, \phi) &=& \delta_{ab} - 2 \int^{\chi_{\ast}}_{0}d\chi\frac{S_k(\chi_\ast-\chi)}{S_k(\chi_{\ast})S_k(\chi)}\nabla_{a}\nabla_{b}\Psi_{W}(t_0-\chi, \chi,\vartheta, \varphi)\, ,\\
& \equiv& \begin{pmatrix}
1-\kappa-\gamma_{1} & -\gamma_{2} \\
-\gamma_{2} & 1-\kappa + \gamma_{1}
\end{pmatrix} \,.
\eea
Here $(a,b)$ are the two angular directions $e_\vartheta$ and $e_\varphi$ and 
\be
\Psi_{W} = \frac{1}{2}(\Phi+\Psi)
\ee
is the mean of the two Bardeen potentials, the so called Weyl potential. The time variable $t$ denotes conformal time. The convergence $\ka$ therefore is
\be\label{e:kappa}
\ka = 1-\frac{1}{2}{\rm trace}\JJ = \int^{\chi_{\ast}}_{0}d\chi\frac{S_k(\chi_\ast-\chi)}{S_k(\chi_{\ast})S_k(\chi)}\nabla^2_\perp\Psi_{W}(t_0-\chi, \chi,\vartheta, \varphi)\,.
\ee
Here $\nabla^2_\perp$ denotes the Laplacian  on the sphere, i.e. wrt $(\vartheta, \varphi)$.

Note that $1+\De(z)$ is not just the mean of the second order perturbation of the determinant of the Jacobi map $\JJ$. The second order corrections to $\langle d_A\rangle$ are much more complicated. They are derived 
in~\cite{BenDayan:2012wi,Fanizza:2013doa,Umeh:2012pn,Umeh:2014ana,Marozzi:2014kua} and $\De(z)$ is the dominant contributions to $\langle d_A\rangle$  at $z\gtrsim 0.2$.

To determine the variance of $\ka$ we consider standard flat $\La$CDM with cosmological parameters from Planck~2018~\cite{Aghanim:2018eyx}. We compute the linear and halofit power spectrum of $\Psi$ with {\sc class}~\cite{class1,class2}. From Eq.~(\ref{e:kappa}) in the case of vanishing curvature we obtain
\bea
\Delta(z_{\ast}) &=& \frac{3}{2}\left\langle\bigg[\int_{0}^{\chi_\ast}\frac{d\chi}{\chi}\bigg(1-\frac{\chi}{\chi_{\ast}}\bigg)\nabla^2_\perp\Psi_W(\chi)\bigg]^{2}\right\rangle
\label{eqn:delint}\\
&=&6\pi\sum_{\ell=0}^{\infty}\bigg[\frac{\ell(\ell+1)}{2\ell+1}\bigg]^{2}\int_{0}^{\chi_{\ast}}d\chi\frac{(\chi_{\ast}-\chi)^{2}}{\chi\chi_{\ast}^{2}}\mathcal{P}_{\Psi}(k,t_0-\chi)\bigg|_{k=(\ell+\frac{1}{2})/\chi} \,.
\label{eqn:delta}
\eea
For the last equation we have used $\Psi_W\simeq\Psi$ and
\be
\Psi(t_0-\chi,\chi,\bn) = \sqrt{\frac{2}{\pi}}\sum_{\ell,m}i^\ell\int d^3k\Psi(t_0-\chi,\bk)j_\ell(k\chi)Y_{\ell m}(\hat \bk)Y^*_{\ell m}(\bn) \,.
\ee
We have then employed the addition theorem of spherical harmonics and the Limber approximation~\cite{Limber:1954zz,LoVerde:2008re} to perform the integrals over $\bk$. $\mathcal{P}_{\Psi}(k,t)$ is the dimensionless power spectrum defined by
\be
k^3\langle \Psi(\bk,t)\Psi(\bk',t)\rangle = 2\pi^2\de(\bk-\bk')\mathcal{P}_{\Psi}(k,t)\,.
\ee
In Appendix~\ref{s:app} we provide a detailed derivation of (\ref{eqn:delta}).

When inserting $\PP_\Psi(k,t)$ from linear perturbation theory (computed with {\sc class}), the sum over $\ell$ in \eqref{eqn:delta} converges quite rapidely. However, when taking into account non-linearities with halofit (also computed with {\sc class}), which is an approximation of the density power spectrum to $N$-body simulations, the sum over $\ell$ first seems to diverge. Only at very high $\ell \sim 10^5$, the power spectrum becomes steeper again so that the sum finally converges. This can be understood analytically, by considering that the halofit density power spectrum on intermediate scales behaves like $P_\de(k)\propto k^\ga$ with $-2<\ga<-1$. Using the Poisson equation this yields for the potential power spectrum $P_\Psi\propto k^{-4}P_\de \propto k^{\ga-4}$ on these scales, hence $\PP_\Psi \propto k^3P_\Psi \propto k^{\ga-1}$. In the sum  \eqref{eqn:delta}  we have to sum roughly $\ell^2\PP_\Psi(\ell/\chi) \propto \ell^{\ga+1}$ which diverges for $\ga>-2$. On linear scales $\ga=-3$ on small scales, which converges nicely.  For the halofit power spectrum, the slope becomes flatter when non-linearities set in, but then turns back to $\ga \simeq -2.3$ on very small scales. We have summed over the flatter part of the spectrum until the assymptotic slope $\ga+1\sim -1.3$ was reached and then integrated to $\ell=\infty$. Even though at very small scales there is probably some smoothing of power due to non-gravitational physics, this is not very relevant as these extremely small scales contribute little to the final result.

\begin{figure}[h!]
	\centering
	\includegraphics[width=0.71\textwidth]{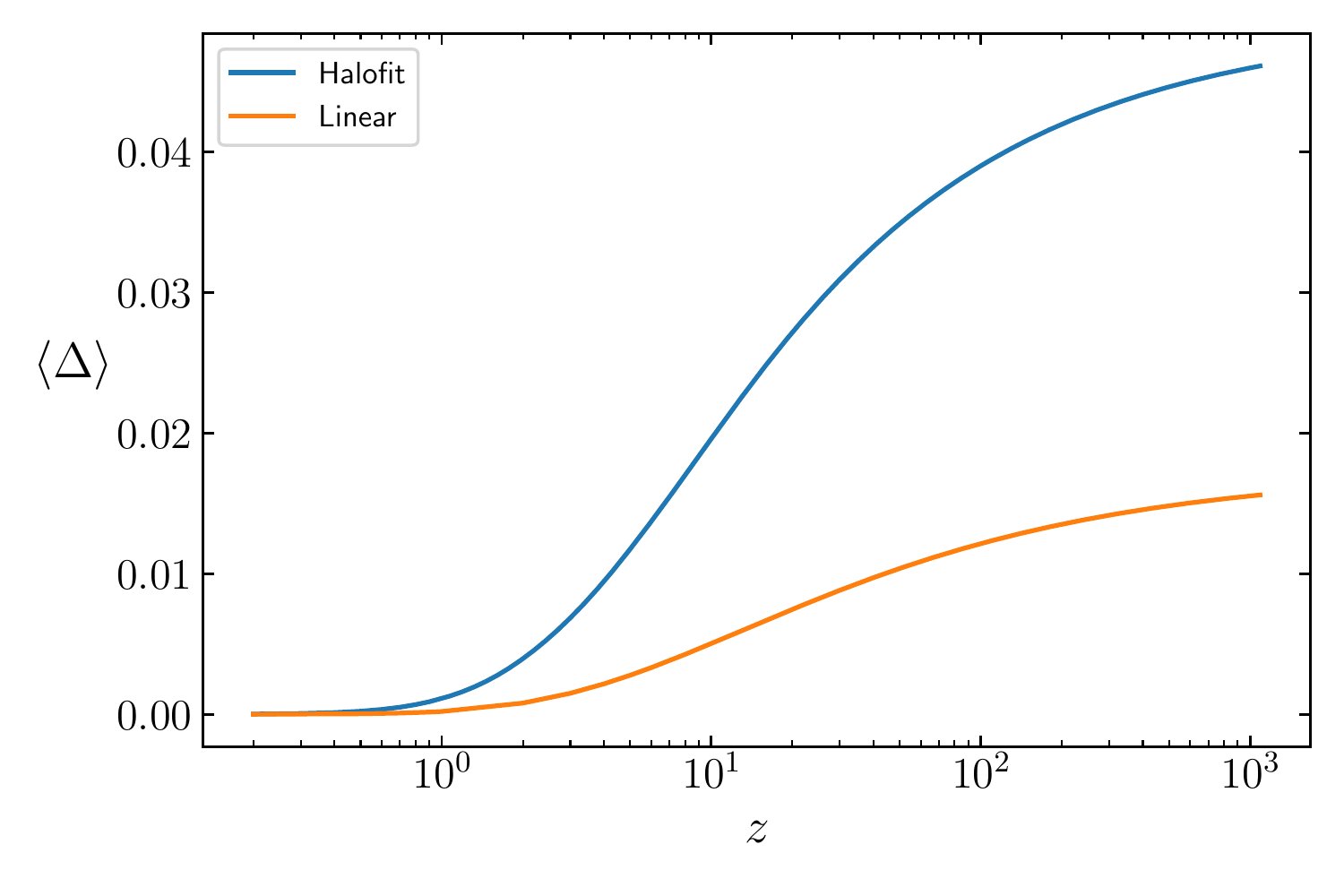}
	\caption{The function $\Delta(z)$ is shown for both, the linear power spectrum and halofit. Note that the halofit result is about three times larger than the one from linear perturbation theory.}
	\label{f:Delta}
\end{figure}

The resulting correction $\De(z)$ from linear perturbation theory and halofit  is shown in Fig.~\ref{f:Delta}.

The correction to the mean distance is about 1\% out to redshift $z\simeq 5$ and grows to about $4$\% by redshift $z>100$ where it flattens to become about 4.6\% at redshift $z\simeq 1000$. It is interesting to note that non-linearities, taken into account here via the halofit model for the matter power spectrum, increase the result by about a factor of 3. Even though non-linearities become irrelevant at $z>5$, since $\De(z)$ is an integrated quantity from redshift 0 out to redshift $z$, nonlinearities are relevant for this correction at all redshifts. However, while the difference between the linear and the non-linear corrections at $z=2$ is about a factor 5 it has reduced to a factor 3 by redshift $z=1000$.

\section{Results}
\label{s:res}

\begin{figure}[hbt!]
	\centering
	\hspace*{-1.05cm} 
	\includegraphics[width=0.7\textwidth]{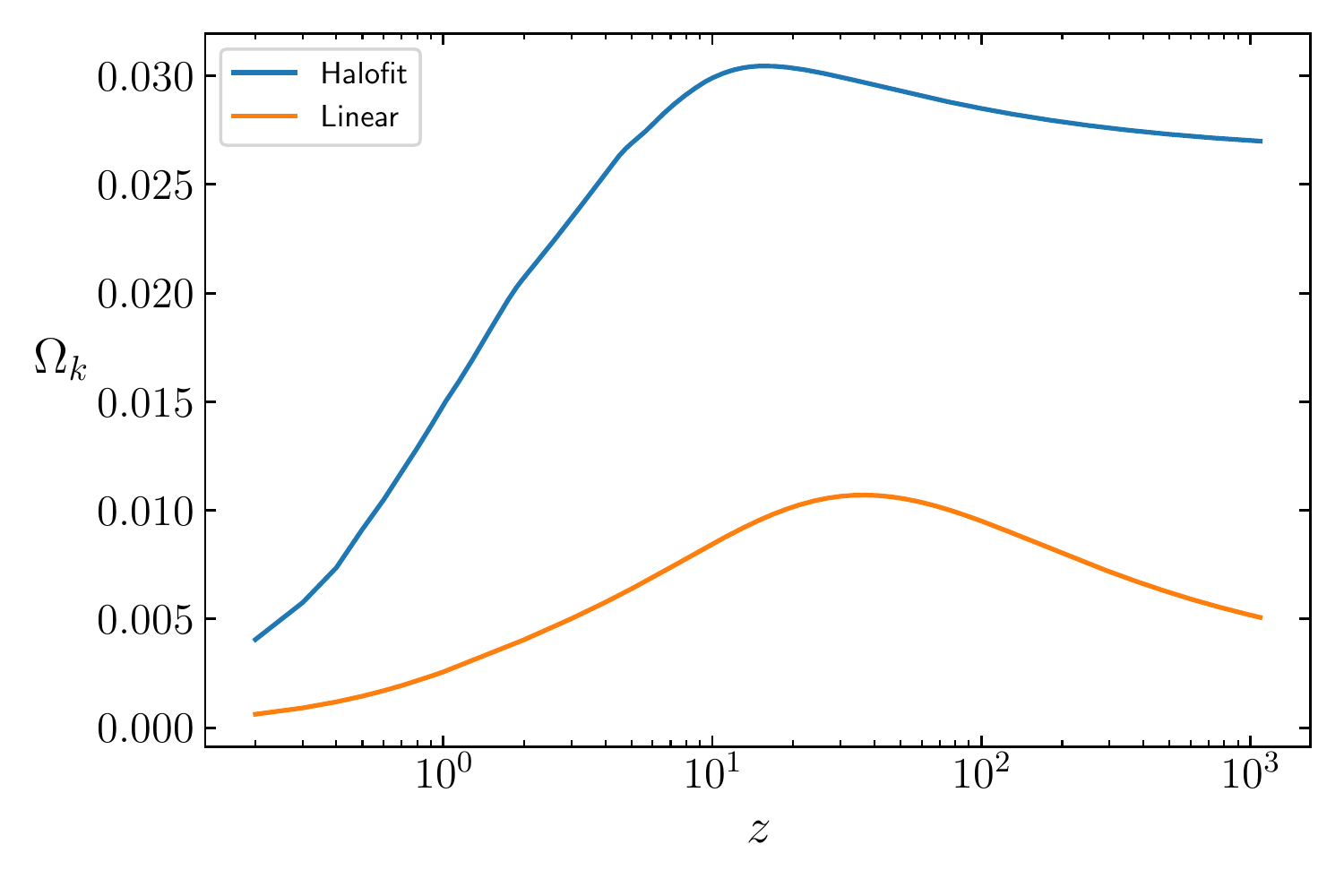}
	\caption{The modified consistency relation is shown for both, the linear power spectrum and halofit.}
	\label{f:cons}
\end{figure}

The correction $\De(z)$ in the angular diameter distance can be used to calculate the modification to the consistency relation (\ref{e:cons}). In a flat Friedmann Universe, the fractional change $\De$ induces an 'apparent' curvature parameter, $\De\Om_k$, which however now depends on redshift. To lowest order in $\De$ we have
\be\label{e:dOmk}
\De\Om_k(z)= 2\frac{D'(z)D(z)H^2(z)\De'(z)+\De(z)}{(D(z)H_0)^2} \,.
\ee
This is shown in Fig.~\ref{f:cons}.

This $\De\Om_k$ induced by structure grows until redshift $z =z_{\max}\sim 11$  ($z_{\max}\sim 30$ in the linear analysis)  while at $z\gtrsim z_{\max}$ it settles nearly to a constant value in the halofit model (slowly decays within linear perturbation theory). The peak height of the non-linear analysis is nearly 3\% which is more than the five times the linear result.  The very significant growth of $\De\Om_k$ at redshifts $0.2\leq z\leq 10$ comes from the steep growth of $\De$ leading to a much larger value of $\De'$ in halofit than within linear perturbation theory. At higher redshifts where linear perturbation theory is valid, the growth of $\De$ becomes actually somewhat slower than the growth of $D$ and the final result decreases slowly. Nevertheless the 'apparent curvature' inferred from non-linear structure at high redshift is about five times larger than the result from linear perturbation theory.

\begin{figure}[hbt!]
	\centering
	\hspace*{-1.05cm} 
	\includegraphics[width=0.7\textwidth]{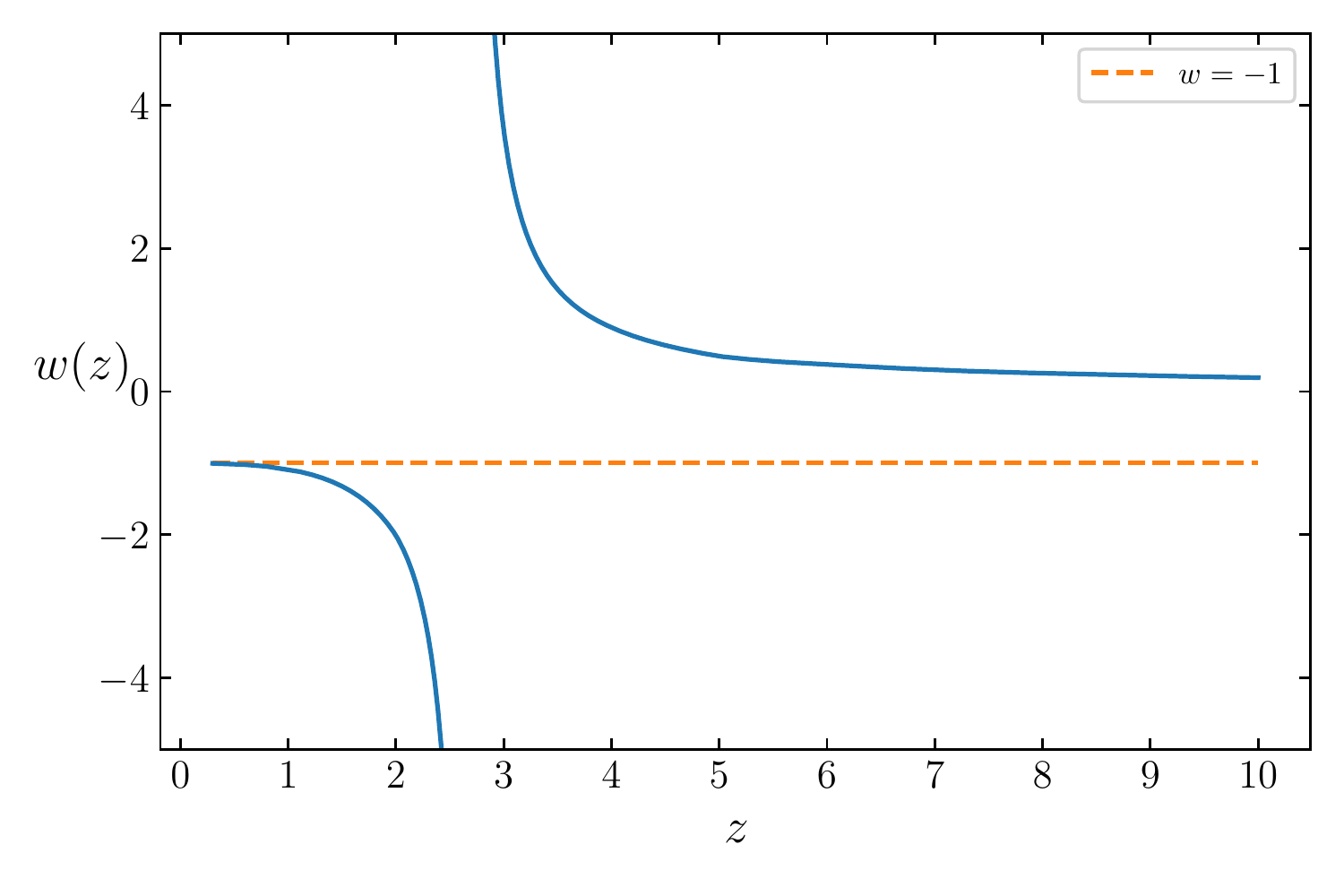}
	\caption{The equation of state $w(z)$ from first and second derivatives if the comoving distance as given in Eq.~\eqref{e:wDz}, in a $\La$CDM Universe with clustering for redshifts $0.2<z<10$.}
	\label{f:state}
\end{figure}

Apart from curvature, one can in principle use distance data to infer the dark energy equation of state, let us denote it by $w(z) = P_{\rm de}(z)/\rho_{\rm de}(z)$.  In a perfect Friedman universe, we can express the function $w(z)$ in terms of the distance $D(z)$ and its first and second derivatives. Assuming that dark energy does not interact with ordinary matter but is separately conserved, dark energy 'conservation' implies
\be\label{e:con-de}
\frac{d\rho_{\rm de}}{dz}(z) = 3\frac{1+w(z)}{1+z}\rho_{\rm de}(z) \quad \Rightarrow \quad
\rho_{\rm de}(z) = \rho_{\rm de}(0)\exp\left[3\int_0^{z}dz'\frac{1+w(z')}{1+z'}\right] \,.
\ee
Since the true \(\Omega_{k}\) of the background universe is very small and compatible with zero, we neglect it in these considerations so that
\be\label{e:w}
D(z) = \int_0^z\frac{dz'}{H(z')} = \frac{1}{H_0} \int_0^z\frac{dz'}{\sqrt{\Om_m(1+z')^3 +\Om_{\rm de}\exp\left[3\int_0^{z'}dz''\frac{1+w(z'')}{1+z''}\right]}} \,.
\ee
After some algebra, we can express $w(z)$ in terms of $\Om_m,~H_0$ as well as first and second derivatives of $D(z)$ (similar expressions can be found in~\cite{Clarkson:2007bc}),
\be\label{e:wDz}
w(z)=\frac{\frac{2}{3}(1+z)\frac{D''}{D'}+1}{(1+z)^3\Om_m(D'H_0)^2-1} \,.
\ee
We want to study how this expression is affected by clustering. 
Replacing $D(z)$ by $D(z)(1+\De(z))$ which is the distance measured in the clustered Universe, we find the behavior of $w(z)$ shown in Fig.~\ref{f:state}.

Of course in practice, it is not possible to take  second derivatives from observational data and workers in the field will not use directly Eq.~\eqref{e:wDz} to determine the dark energy equation of state from data. One will have to use fitting functions for $D(z)$ and study how the inferred $w(z)$ depends on the fitting parameters. But if successful, determining the effects of clustering on $w(z)$ might be a promising tool to study, e.g.,  the onset of non-linear clustering.

At small redshifts, $z<1$, the deviation from $w=-1$ induced by clustering is small but always  in the direction to make $w<-1$.  At $z>1.5$ the deviation becomes large and it even diverges at $z\sim 2.5$. At this redshift $(1+z)^3\Om_m(H_0[D(z)(1+\De(z))]')^2 =1$ such that the denominator of \eqref{e:wDz} vanishes in the clustered Universe. At even higher redshifts, $w(z)>0$ and tends to $0$ from above. These very significant effects from clustering are in agreement with the very significant effects from neglecting a possible curvature which have been discussed in Ref.~\cite{Clarkson:2007bc}.

\section{Conclusions}
\label{s:con}
In this paper we have studied the effect of clustering in the cosmological standard model on the consistency relations for $\Om_k$ and $w(z)$  introduced in Refs.~\cite{Clarkson:2007bc,Clarkson:2007pz}. While the relation \eqref{e:cons}  holds roughly at the 1\% level below $z\simeq 1$ in the clustered Universe, we expect it to deviate by more than 2\% at redshift $z\gtrsim 5$. As present Planck limits on spatial curvature are already significantly better than these values~\cite{Aghanim:2018eyx}, we conclude that the consistency test is not very useful to measure spatial curvature. Nevertheless, it remains an important consistency test as a diagnostic for systematic errors in measurements of cosmological distances and of the Hubble parameter $H(z)$ as they will be possible in the near future with Alcock-Paczynski type measurements e.g. of the BAOs (baryon acoustic oscillations)~\cite{Reid:2012,Blake:2012pj,Montanari:2012me,Lepori:2016rzi}. Furthermore, at relatively low redshift, $0.2<z<2$ the inferred value $\Om_k$ is very small and, when accounting for the correction from clustering, one may achieve an accuracy comparable to the Planck value from supernova data at intermediate redshift.

On the other hand, the deviation of this test due to clustering will be a good measure of the integrated matter power spectrum and of the variance of the integrated convergence, $\langle\ka^2\rangle (z)$.
A deviation of this test by more then a few percent would be a strong indication of either systematic problems in  distance or in $H(z)$ measurements, or that our Universe cannot be described as Friedmann Universe with small fluctuations which only relatively recently have grown into non-linear structures.

Alternatively, measuring $D(z)$ and $\Om_m,~H_0$ independently, one can use these to measure the dark energy equation of state, $w(z)$ using Eq.~\eqref{e:wDz}. We have  shown that this measurement is  very strongly affected by clustering for redshifts $z>1.5$, but can provide a useful constraint at relatively low redshifts, $0.2<z<0.7$.  The divergence at $z\sim 2.5$ can be used as a very sensitive measure of the overall amplitude of clustering. 

Of course $D(z)$ as obtained from data is very noisy and it will not be possible to take second derivatives directly from the data. It will be necessary to first smooth the data or to model it with a fitting function. In Ref.~\cite{Clarkson:2007bc} it has also been shown that the reconstructed equation of state, $w(z)$ is very sensitive to our assumptions on curvature. A 1\% error on curvature results in a 1\% error on $w(z)$ out to redshift $z\sim 2$. At higher redshift, the required accuracy on curvature becomes more and more demanding. 
Similarly, as we show in Fig.~\ref{f:state}, at redshifts $z>1.5$ the $w(z)$ inferred from measured distances in the clustered Universe is very different from $-1$ even if the true dark energy is simply a cosmological constant.

\section*{Acknowledgements}
We thank Chris Clarkson for interesting comments and the anonymous Referee for pointing out an error in our numerical treatment in a first version. This work is supported by the Swiss National Science Foundation.

\appendix
\section{The variance of the convergence $\ka$}\label{s:app}
In this appendix we give some details on the computation of $\De(z)=(3/2)\langle\ka^2\rangle(z)$. According to \eqref{e:kappa}, in a spatially flat universe, $K=0$, and neglecting anisotropic stress so that $\Psi_W=\Psi$,  we have
\be\label{e:kaA}
\ka(z_*) =  \int^{\chi_{\ast}}_{0}d\chi\frac{(\chi_\ast-\chi)}{\chi_{\ast}\chi}\nabla^2_\perp\Psi(t_0-\chi, \chi\bn)\,,
\ee
where $\chi_*=\chi(z_*)$ and $\nabla_\perp$ is the (dimensionless) angular gradient on the sphere.

We represent $\Psi$ by its Fourier transform (careful, {\sc class} uses unitary Fourier convention!).
Denoting $\Psi(t_0-\chi, \chi\bn) \equiv \Psi(\chi,\bn)$ we have
\be
\Psi(\chi,\bn) = \frac{1}{(2\pi)^{3/2}}\int d^3k e^{i\bk\cdot\bn\chi}\hat\Psi(t_0-\chi,\bk) \,.
\ee
We use
$$
e^{i\bk\cdot\bn\chi} = 4\pi\sum_{\ell,m}i^\ell Y_{\ell m}(\hat\bk) Y_{\ell m}^*(\bn)j_\ell(k\chi)\,,
$$
where $Y_{\ell m}$ are spherical harmonics, $j_\ell$ are spherical Bessel functions, $\hat\bk=\bk/k$ and a star denotes complex conjugation. Using that  $\nabla^2_\perp Y_{\ell m}^*(\bn)=-\ell(\ell+1) Y_{\ell m}^*(\bn)$, we can write
\bea
\Psi(\chi,\bn) &=& \frac{4\pi}{(2\pi)^{3/2}}\sum_{\ell,m}i^\ell Y_{\ell m}^*(\bn)\int d^3k  Y_{\ell m}(\hat
\bk)j_\ell(k\chi)\hat\Psi(t_0-\chi,\bk) \\
\nabla^2_\perp\Psi(\chi,\bn) &=& \frac{-4\pi}{(2\pi)^{3/2}}\sum_{\ell,m}i^\ell\ell(\ell+1) Y_{\ell m}^*(\bn)\int d^3k  Y_{\ell m}(\hat\bk)j_\ell(k\chi)\hat\Psi(t_0-\chi,\bk)\,.\label{e:LapPsiF}
\eea
Since the left hand side of \eqref{e:LapPsiF} is real, this equation remains correct when we take the complex conjugate of all  terms on the right hand side. We now want to use this in
\be
\langle\ka^2\rangle(z_*)=\int^{\chi_{\ast}}_{0}d\chi\int^{\chi_{\ast}}_{0}d\chi'\frac{(\chi_\ast-\chi)(\chi_\ast-\chi')}{\chi^2_{\ast}\chi\chi'}\langle\nabla^2_\perp\Psi_(\chi,\bn)\nabla^2_\perp\Psi_(\chi',\bn)\rangle\,.
\ee
We insert \eqref{e:LapPsiF}  and its complex conjugate for $\nabla^2_\perp\Psi_(\chi,\bn)$ and $\nabla^2_\perp\Psi_(\chi',\bn)$, using the definition of the dimensionless power spectrum $\PP_\Psi$,
\be
\langle\Psi(t,\bk)\Psi^*(t',\bk')\rangle = \frac{2\pi^2}{k^3}\de(\bk-\bk')\PP_\Psi(k,t,t') \,.
\ee
The Dirac--delta removes the $d^3k'$ integration. The angular integration over $\hat\bk$ only concerns the spherical harmonics and yields
$$
\int d\Om_{\hat\bk}Y_{\ell m}(\hat\bk)Y^*_{\ell' m'}(\hat\bk) = \de_{\ell\ell'}\de_{mm'} \,.
$$
This removes the double sum over $\ell'$ and $m'$ so that we obtain
\bea
\langle\ka^2\rangle(z_*) &=& 4\pi\sum_{\ell m}\left[\ell(\ell+1)\right]^2\left|Y_{\ell m}(\bn)\right|^2\int^{\chi_{\ast}}_{0}d\chi\int^{\chi_{\ast}}_{0}d\chi'\frac{(\chi_\ast-\chi)(\chi_\ast-\chi')}{\chi^2_{\ast}\chi\chi'} \nonumber\\
&& \hspace{2cm} \times\int \frac{dk}{k}\PP_\Psi(k,t_0-\chi,t_0-\chi')j_\ell(k\chi)j_\ell(k\chi') \,.
\eea
The sum over $m$ can now simply be performed using
hat
$$
\sum_m \left|Y_{\ell m}(\bn)\right|^2 =\frac{2\ell+1}{4\pi} \,.
$$
To simplify the expression further we apply Limber's approximation~\cite{Limber:1954zz,LoVerde:2008re} which claims that for a sufficiently smooth/flat function $f$ with converging integral we have
\be\label{e:Limber}
\int\frac{dk}{k}f(k)j_\ell(k\chi)j_\ell(k\chi') =\frac{4\pi}{(2\ell+1)^3}\chi\de(\chi-\chi')f\left(\frac{\ell+1/2}{\chi}\right) \,.
\ee
This has been tested for the lensing potential and has been found to be a good approximation.
It solves the integral over $k$ and the Dirac--delta removes the integral over $\chi'$. We then finally obtain
\be
\langle\ka^2\rangle(z_*)=4\pi\sum_{\ell}\left[\frac{\ell(\ell+1)}{2\ell+1}\right]^2 \int^{\chi_{\ast}}_{0}d\chi\frac{(\chi_\ast-\chi)^2}{\chi^2_{\ast}\chi}\PP_\Psi\left(\frac{\ell+1/2}{\chi},t_0-\chi\right) \,.
\ee
The relation $\De =3/2\langle\ka^2\rangle$ now implies Eq.~\eqref{eqn:delta}.


\begin{thebibliography}{10}

\bibitem{Clarkson:2007bc}
C.~Clarkson, M.~Cortes, and B.~A. Bassett, {\it {Dynamical Dark Energy or
  Simply Cosmic Curvature?}},  {\em JCAP} {\bf 0708} (2007) 011,
  [\href{http://arxiv.org/abs/astro-ph/0702670}{\tt astro-ph/0702670}].

\bibitem{Clarkson:2007pz}
C.~Clarkson, B.~Bassett, and T.~H.-C. Lu, {\it {A general test of the
  Copernican Principle}},  {\em Phys. Rev. Lett.} {\bf 101} (2008) 011301,
  [\href{http://arxiv.org/abs/0712.3457}{{\tt arXiv:0712.3457}}].

\bibitem{BenDayan:2012wi}
I.~Ben-Dayan, G.~Marozzi, F.~Nugier and G.~Veneziano,
{\it The second-order luminosity-redshift relation in a generic inhomogeneous cosmology},
JCAP {\bf 1211}, 045 (2012), doi:10.1088/1475-7516/2012/11/045, [\href{http://arxiv.org/abs/1209.4326}{{\tt arXiv:1209.4326}} [astro-ph.CO]].

\bibitem{Fanizza:2013doa}
G.~Fanizza, M.~Gasperini, G.~Marozzi and G.~Veneziano,
{\it An exact Jacobi map in the geodesic light-cone gauge},
JCAP {\bf 1311}, 019 (2013)
doi:10.1088/1475-7516/2013/11/019 [\href{http://arxiv.org/abs/1308.4935}{{\tt arXiv:1308.4935}} [astro-ph.CO]].

\bibitem{Umeh:2012pn}
O.~Umeh, C.~Clarkson, and R.~Maartens, {\it {Nonlinear relativistic corrections
  to cosmological distances, redshift and gravitational lensing magnification:
  I. Key results}},  {\em Class. Quant. Grav.} {\bf 31} (2014) 202001,
  [\href{http://arxiv.org/abs/1207.2109}{{\tt arXiv:1207.2109}}].

\bibitem{Umeh:2014ana}
O.~Umeh, C.~Clarkson, and R.~Maartens, {\it {Nonlinear relativistic corrections
  to cosmological distances, redshift and gravitational lensing magnification.
  II - Derivation}},  {\em Class. Quant. Grav.} {\bf 31} (2014) 205001,
  [\href{http://arxiv.org/abs/1402.1933}{{\tt arXiv:1402.1933}}].

\bibitem{Marozzi:2014kua}
G.~Marozzi, {\it {The luminosity distance–redshift relation up to second
  order in the Poisson gauge with anisotropic stress}},  {\em Class. Quant.
  Grav.} {\bf 32} (2015), no.~4 045004,
  [\href{http://arxiv.org/abs/1406.1135}{{\tt arXiv:1406.1135}}]. [erratum:
  Class. Quant. Grav.32,179501(2015)].

\bibitem{Clarkson:2014pda}
C.~Clarkson, O.~Umeh, R.~Maartens, and R.~Durrer, {\it {What is the distance to
  the CMB?}},  {\em JCAP} {\bf 1411} (2014), no.~11 036,
  [\href{http://arxiv.org/abs/1405.7860}{{\tt arXiv:1405.7860}}].

\bibitem{Bonvin:2005ps}
C.~Bonvin, R.~Durrer, and M.~A. Gasparini, {\it {Fluctuations of the luminosity
  distance}},  {\em Phys. Rev.} {\bf D73} (2006) 023523,
  [\href{http://arxiv.org/abs/astro-ph/0511183}{{\tt astro-ph/0511183}}].
  [Erratum: Phys. Rev.D85,029901(2012)].

\bibitem{BenDayan:2013gc}
I.~Ben-Dayan, M.~Gasperini, G.~Marozzi, F.~Nugier, and G.~Veneziano, {\it
  {Average and dispersion of the luminosity-redshift relation in the
  concordance model}},  {\em JCAP} {\bf 1306} (2013) 002,
  [\href{http://arxiv.org/abs/1302.0740}{{\tt arXiv:1302.0740}}].

\bibitem{Perlick:2004tq}
V.~Perlick, {\it {Gravitational lensing from a spacetime perspective}},  {\em
  Living Rev. Rel.} {\bf 7} (2004) 9.

\bibitem{RuthBook}
R.~Durrer, {\em The Cosmic Microwave Background}.
\newblock Cambridge University Press, 2008.

\bibitem{Aghanim:2018eyx}
{\bf Planck} Collaboration, N.~Aghanim et~al., {\it {Planck 2018 results. VI.
  Cosmological parameters}},  \href{http://arxiv.org/abs/1807.06209}{{\tt
  arXiv:1807.06209}}.

\bibitem{class1}
J.~{Lesgourgues}, {\it {The Cosmic Linear Anisotropy Solving System (CLASS) I:
  Overview}},  {\em ArXiv e-prints} (Apr., 2011)
  [\href{http://arxiv.org/abs/1104.2932}{{\tt arXiv:1104.2932}}].

\bibitem{class2}
D.~Blas, J.~Lesgourgues, and T.~Tram, {\it {The Cosmic Linear Anisotropy
  Solving System (CLASS) II: Approximation schemes}},  {\em JCAP} {\bf 1107}
  (2011) 034, [\href{http://arxiv.org/abs/1104.2933}{{\tt arXiv:1104.2933}}].

\bibitem{Limber:1954zz}
D.~N. Limber, {\it {The Analysis of Counts of the Extragalactic Nebulae in
  Terms of a Fluctuating Density Field. II}},  {\em Astrophys. J.} {\bf 119}
  (1954) 655.

\bibitem{LoVerde:2008re}
M.~LoVerde and N.~Afshordi, {\it {Extended Limber Approximation}},  {\em Phys.
  Rev.} {\bf D78} (2008) 123506, [\href{http://arxiv.org/abs/0809.5112}{{\tt
  arXiv:0809.5112}}].

\bibitem{Reid:2012}
B.~A. Reid et~al., {\it {The clustering of galaxies in the SDSS-III Baryon
  Oscillation Spectroscopic Survey: measurements of the growth of structure and
  expansion rate at z = 0.57 from anisotropic clustering}},  {\em Mon. Not.
  Roy. Ast. Soc.} {\bf 426} (Nov, 2012) 2719--2737,
  [\href{http://arxiv.org/abs/1203.6641}{{\tt arXiv:1203.6641}}].

\bibitem{Blake:2012pj}
C.~Blake et~al., {\it {The WiggleZ Dark Energy Survey: Joint measurements of
  the expansion and growth history at z < 1}},  {\em Mon. Not. Roy. Astron.
  Soc.} {\bf 425} (2012) 405--414, [\href{http://arxiv.org/abs/1204.3674}{{\tt
  arXiv:1204.3674}}].

\bibitem{Montanari:2012me}
F.~Montanari and R.~Durrer, {\it {A new method for the Alcock-Paczynski test}},
   {\em Phys. Rev.} {\bf D86} (2012) 063503,
  [\href{http://arxiv.org/abs/1206.3545}{{\tt arXiv:1206.3545}}].

\bibitem{Lepori:2016rzi}
F.~Lepori, E.~Di~Dio, M.~Viel, C.~Baccigalupi, and R.~Durrer, {\it {The Alcock
  Paczynski test with Baryon Acoustic Oscillations: systematic effects for
  future surveys}},  {\em JCAP} {\bf 1702} (2017), no.~02 020,
  [\href{http://arxiv.org/abs/1606.03114}{{\tt arXiv:1606.03114}}].

\end{thebibliography}
\end{document}